\documentclass[9pt,twocolumn,twoside]{JournCls}
\usepackage{scalerel}
\usepackage{tikz}
\usetikzlibrary{svg.path,calc}

\definecolor{orcidlogocol}{HTML}{A6CE39}
\tikzset{
   orcidlogo/.pic={
    \fill[orcidlogocol] svg{M256,128c0,70.7-57.3,128-128,128C57.3,256,0,198.7,0,128C0,57.3,57.3,0,128,0C198.7,0,256,57.3,256,128z};
    \fill[white] svg{M86.3,186.2H70.9V79.1h15.4v48.4V186.2z}
                 svg{M108.9,79.1h41.6c39.6,0,57,28.3,57,53.6c0,27.5-21.5,53.6-56.8,53.6h-41.8V79.1z M124.3,172.4h24.5c34.9,0,42.9-26.5,42.9-39.7c0-21.5-13.7-39.7-43.7-39.7h-23.7V172.4z}
                 svg{M88.7,56.8c0,5.5-4.5,10.1-10.1,10.1c-5.6,0-10.1-4.6-10.1-10.1c0-5.6,4.5-10.1,10.1-10.1C84.2,46.7,88.7,51.3,88.7,56.8z};
  }
}

\newcommand\orcidicon[1]{\href{https://orcid.org/#1}{\mbox{
\begin{tikzpicture}[overlay,remember picture]
\coordinate (A);
\coordinate(B) at ($(A)-(2pt,-9pt)$);
\end{tikzpicture}
\begin{tikzpicture}[overlay,remember picture,yscale=-0.045,xscale=0.045,transform shape]
\pic at (B) {orcidlogo};
\end{tikzpicture}
}{}}}

\usepackage{hyperref} 
 
\usepackage{caption}
\usepackage{balance}
\DeclareCaptionFormat{myformat}{#1#2#3\hrulefill}
\journal{JournSty} 
\setboolean{shortarticle}{false} 
\title{Directional Supercontinuum Generation: The Role of the Soliton}

\author[1,*,\protect\orcidicon{0000-0002-0139-1141}\,\,]{Simon Christensen}
\author[1,\protect\orcidicon{0000-0003-2007-0930}\,\,]{ Shreesha Rao D. S.}
\author[1,2,\protect\orcidicon{0000-0002-8041-9156}\,\,]{Ole Bang}
\author[1,\protect\orcidicon{0000-0002-9407-9236}\,\,]{Morten Bache}
\affil[1]{DTU Fotonik, Department of Photonics Engineering, Technical University of Denmark, \O rsteds Plads, 2800 Kongens Lyngby, Denmark.}
\affil[2]{NKT Photonics A/S, Blokken 84, 3460 Birker\o d, Denmark.}
\affil[*]{Corresponding author: sichr@fotonik.dtu.dk}

\begin{abstract}
In this paper we numerically study supercontinuum generation by pumping a silicon nitride waveguide, with two zero-dispersion wavelengths, with femtosecond pulses. The waveguide dispersion is designed so that the pump pulse is in the normal-dispersion regime. We show that because of self-phase modulation, the initial pulse broadens into the anomalous-dispersion regime, which is sandwiched between the two normal-dispersion regimes, and here a soliton is formed. The interaction of the soliton and the broadened pulse in the normal-dispersion regime causes additional spectral broadening through formation of dispersive waves by non-degenerate four-wave mixing and cross-phase modulation. This broadening occurs mainly towards the second normal-dispersion regime. We show that pumping in either normal-dispersion regime allows broadening towards the other normal-dispersion regime. This ability to steer the continuum extension towards the direction of the other normal-dispersion regime, beyond the sandwiched anomalous-dispersion regime underlies the directional supercontinuum notation. We numerically confirm the approach in a standard silica microstructured fiber geometry with two zero-dispersion wavelengths.
\end{abstract}
\setboolean{displaycopyright}{true}
\begin{document}
\maketitle
\section{Introduction}
In the last two decades research in supercontinuum generation (SCG) has gained a massive amount of interest; especially the use of highly confining waveguides has revolutionized the field. The two major platforms in play are microstructured fibers (MSF) and waveguides in nonlinear materials. Here we focus on SCG using femtosecond (fs) pump pulses because it can provide very low noise~\cite{Dudley02C,IvnRIN}. Two key regions are important when considering group-velocity dispersion ($\beta_2$) of the waveguide: the anomalous-dispersion region (ADR), $\beta_2<0$ and the normal-dispersion region (NDR), $\beta_2>0$. SCG with fs pulses in the NDR is caused by self-phase modulation (SPM) in the early stage and so-called optical wave-breaking in the later stage, a process caused by an interplay of chromatic dispersion and four-wave mixing (FWM) between the newly generated spectral components and the undepleted pump. This results in a simple spectral and temporal shape~\cite{Finot08Ben,HeidtFlatTop10}. SCG with fs pulses in the ADR is governed by solitons, whose interactions lead to a large bandwidth SC, but often a complex spectral and temporal shape. Dispersive waves (DWs) are one of the key effects when generating SC in the ADR~\cite{Akh95DWInt,Sch05Sin,PhasMat04Gen,Pet17TwoDisW}. The DWs are generated by degenerate FWM of the soliton photons, and are generally phase matched outside the ADR. This significantly extends the SC bandwidth. The DW part of a supercontinuum (SC) has been used in many applications, such as confocal microscopy~\cite{Shi04ConMic}, frequency combs~\cite{Skryabin:17}, lidar~\cite{Trocha887} and spectroscopy~\cite{Dutte1701858}. 

A general quest is to broaden the SC, perhaps even towards non-traditional wavelength regimes. There has been significant effort in using the interaction between two pump pulses or solitons~\cite{Sch88SupRSF,Gen05Rou}, and between a soliton and a dispersive wave~\cite{Gen04Eff,Dem14CohSup,Liu15MidProb,InterSolDWEfi05}. Two pulse collision by means of a soliton and a dispersive wave was theoretically shown to produce a coherent SC, which enables the temporal compression of the pulse. This was achieved by using cross-phase modulation (XPM) between waves as the broadening mechanism, therefore avoiding soliton fission and modulation stability~\cite{Dem13Com}. The effect of XPM on Raman solitons was investigated in Ref.~\cite{Roy:11}, and the shift was shown to depend on the dispersion slope of the fiber. Two-color pulse collisions were also studied as event horizon analogies~\cite{Web14NlEve,Phi08Eve,Gu15Pob} and for making  an all-optical transistor~\cite{Dem11Con}. A key weakness of the two-color schemes is the need for two input pulses, which complicates them. Here we show that using a single-pump pulse in the NDR leads to a soliton in the ADR. The interaction of the NDR pulse and the ADR soliton share great similarity with the two-color studies. A similar idea based on an input pulse of dark solitons was investigated in Ref.~\cite{Milian:17}. An interesting prospect would be to use DWs to enhance the SC bandwidth and extend it into new regions when pumping in the NDR. While it was shown to be possible to generate DWs even without solitons~{\cite{Webb13NormalDispDW,rev1_Sun:10,rev1_Phys1,rev1_Phys2}, the generated DWs are extremely weak. However, an interesting recent experiment used a silicon nitride waveguide with two zero-dispersion wavelengths (ZDW)~\cite{Oka17Coh}: here the ADR is sandwiched between two NDRs. When pumped in the IR NDR, the authors found a cascade process eventually generated DWs in the visible NDR. It was also shown that the generated spectrum is coherent. The cascade process was explained by considering dispersive waves generated by both the initial pulse and a compressed dispersive wave in the ADR. Here we investigate this scheme further and show that wave interaction in the ADR plays an important role. The starting point was silicon-rich nitride (SiRN) waveguides with two ZDWs. In such a waveguide one of us recently experimentally observed a continuum generating when pumping in the IR NDR~\cite{Liu16SiRN}. Numerical simulations (not published) indeed indicated a similar behavior as later found in~\cite{Oka17Coh}, namely a directional extension of the SC into the second NDR in the visible. This coins the phrase directional SCG, and in this work we use numerical simulations to investigate the nonlinear processes behind directional SCG. We find that the edge of the SPM broadened pump pulse in the NDR will leak into the ADR, in which solitons can form. The soliton will then be repulsed from the ZDW and move across the ADR, which happens due to XPM from the pulse in the NDR. It is the degenerate FWM from the soliton and non-degenerate FWM between the soliton and the continuum in the pump-NDR that will lead to the formation of DWs with significant spectral densities in the second NDR. By choosing to pump in the NDR located either at shorter or longer wavelength than the ADR, one can ``direct'' the SCG process towards the other NDR. 

We show that by tailoring the SiRN waveguide-dispersion profile a 1.3 octave SC spanning from 0.750 $\mu$m to 1.85 $\mu$m can be generated by using a 1.56 $\mu$m femtosecond pump laser. Importantly, the same waveguide generates a SC from 0.750 $\mu$m to 2.0 $\mu$m when pumped at 0.94 $\mu$m, i.e. when the SCG process is directed towards the longer wavelengths. The similar bandwidth with very different input wavelengths clearly demonstrates the directional broadening of the scheme. It is shown that the chosen dispersion profile allows for the generation of two pulse interactions by generating a soliton in the ADR, even when pumped in the NDR. To show the generality of the scheme a silica MSF geometry is simulated and nonlinear interactions similar to that in the waveguide is observed. As the spectrum mainly broadens through the ADR, the direction of the spectral broadening can be controlled by either changing the pump wavelength or the dispersion profile of the waveguide. The tunability of the spectrum allows the generation of broadband spectrum in the spectral area needed for a specific application. 

\section{Numerical simulations}
The directional SCG process was studied by numerically solving the generalized nonlinear Schr\"{o}dinger equation (GNLSE)~\cite{Agr12NonFib}. The time domain GNLSE with  electric field envelope $A(z,T)$ and $\tilde{A}(z,\omega$) is transformed to frequency domain interaction picture (IP)~\cite{Hult07}. The electric field envelope in the IP is denoted by $C(z,T)$ and $\tilde{C}(z,\omega$). The relation to $\tilde{A}(z,\omega$) is  $\tilde{C}(z,\omega)=~[A_{eff}(\omega)/A_{eff}(\omega_0)]^{-\frac{1}{4}}\tilde{A}(z,\omega)$. Here $z$ is the propagation distance, $\omega$ is the angular frequency, $\omega_0$ is the angular frequency at the pump and $A_{eff}$ is the effective area. 

The GNLSE in the IP can be written as follows in Eq.~(\ref{eq:IPE})~\cite{Laeg07Mode,Dudl10SupGen}:
\begin{multline} \label{eq:IPE}
\frac{\partial\tilde{C}}{\partial z}-i\{\beta(\omega)-[\beta(\omega_0)+\beta_1(\omega_0)(\omega-\omega_0)]\}\tilde{C}+\frac{\alpha(\omega)}{2}\tilde{C}\\=i\bar{\gamma}(\omega)\Big[1+\frac{\omega-\omega_0}{\omega_0}\Big]\mathcal{F}\Big\{C\int_{-\infty}^{\infty}R(T')\mid C(T-T')\mid^2dT'\Big\}.
\end{multline}
Here $T$ is the time in the frame co-moving with the pump, $\mathcal{F}$ denotes the Fourier transform operator, $\beta(\omega)$ is the propagation constant, $\beta_1$ is the inverse pump group-velocity, $\alpha(\omega)$ is the linear loss coefficient, $\bar{\gamma}(\omega)$ is the modified nonlinear coefficient all as a function of frequency, as follows:
\begin{equation}
\bar{\gamma}(\omega)=\frac{n_2n_0\omega_0}{cn_{eff}(\omega)\sqrt{A_{eff}(\omega)A_{eff}(\omega_0)}}, \label{eq:GBar}
\end{equation} 
\begin{equation} \label{eq:Aeff}
A_{eff}(\omega)=\frac{\Big(\int\int_{-\infty}^{+\infty}|F(x,y,\omega)|^{2}dxdy\Big)^{2}}{\int\int_{-\infty}^{+\infty}|F(x,y,\omega)|^{4}dxdy}.
\end{equation}
In the case of the waveguide, the nonlinear coefficient has to include the vectorial components, and it is calculated following \cite{AfsharV09}. For the fiber the definition in Eq.~(\ref{eq:Aeff}) is used. Note that when the effective area ($A_{eff}$) and the effective refractive index ($n_{eff}$) are independent of the frequency, $\bar{\gamma}(\omega)$ becomes the usual nonlinear coefficient $\gamma=\tfrac{n_2\omega_0}{cA_{eff}(\omega_0)}$. $n_2$ is the nonlinear refractive index of the material, $n_0$=$n_{eff}(\omega_0)$, and c is the speed of light in free space. $R(T)$ is the material response function, with a single Lorentzian lineshape model, $R(T)=(1-f_{R})\delta(T)+f_{R}(\tau_{1}^{-2}+\tau_{2}^{-2})\tau_{1}\exp(- T/\tau_{2})\sin(T/\tau_{1})$, where $f_{R}$ is the fractional contribution of the delayed Raman response $\tau_{1}$ is the Raman oscillation period, and $\tau_{2}$ is the Raman decay time.  The  pulse evolution is found by solving Eq.~(\ref{eq:IPE}) using an ordinary differential equation solver.

\subsection{Silicon-rich nitride waveguide}
The waveguide considered here is a 900 by 650 nm SiRN waveguide using the composition characterized by a DCS:NH$_3$ gas ratio of 3.9~\cite{Kruckel}. The SiRN waveguide has a $\gamma$ of 2.9 W$^{-1}$m$^{-1}$ at 1.56 $\mu$m. The Raman effect was also modeled~\cite{Agr12NonFib}. We use a single Lorentzian lineshape model with $f_{R}=0.2$, $\tau_{1}$=13 fs, and $\tau_{2}$=150 fs \cite{Liu16SiRN}. The dispersion profile of the waveguide is optimized to ensure that the 1.56 $\mu$m pump laser is in the NDR, but close to the ADR. The waveguide and dispersion profile of interest are shown in Fig.~\ref{fig:WGADisp}.  
\begin{figure}[t]
\includegraphics[width=0.48\textwidth]{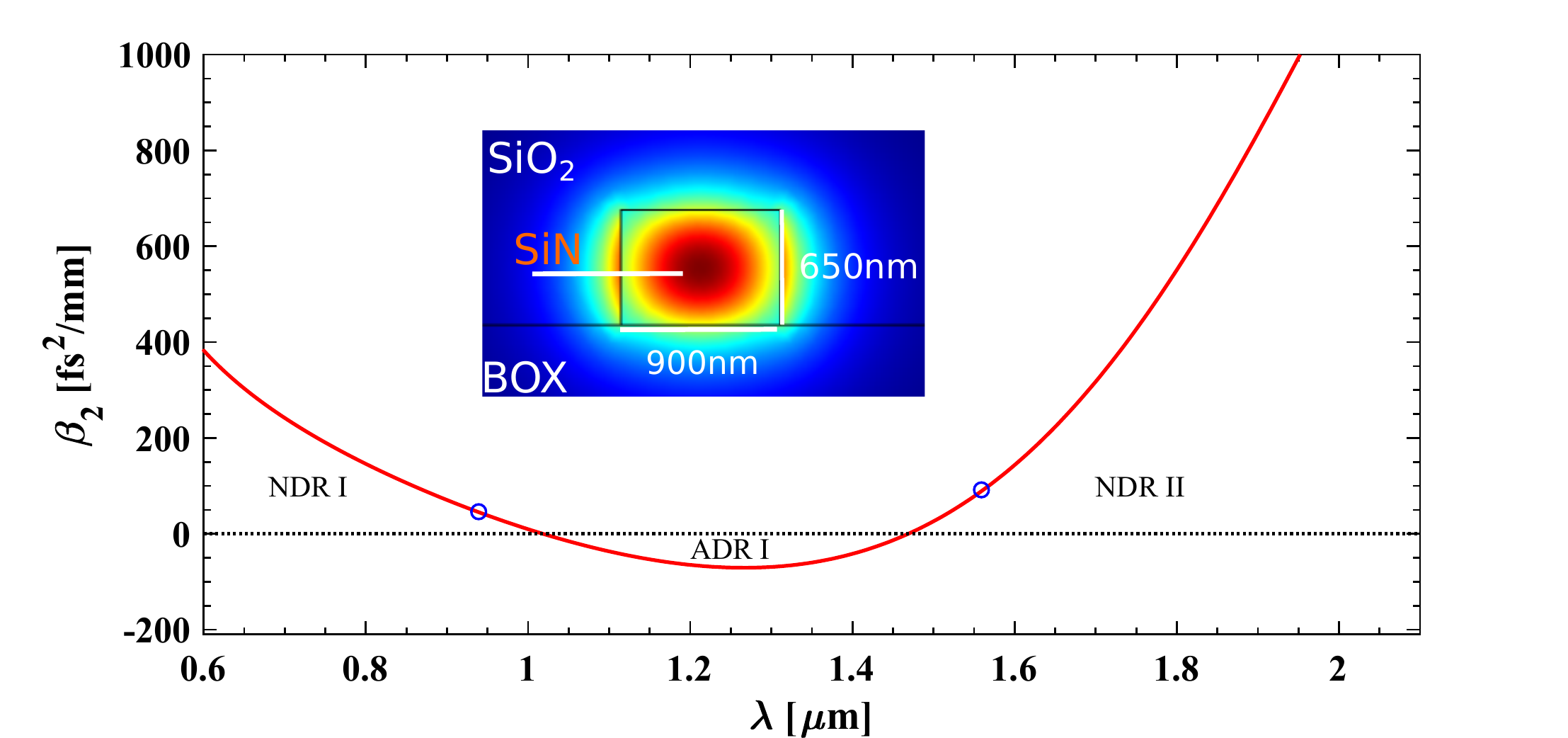}
\caption{GVD ($\beta_2$) for the SiRN waveguide. The circles mark the pump wavelengths used in the paper. The inset represents the COMSOL simulation of the waveguide and the mode obtained.}
\label{fig:WGADisp}
\end{figure}
The modes were found by using COMSOL Multiphysics. The ZDWs are located at 1.02 $\mu$m and 1.47 $\mu$m. For the sake of convenience, the wavelength region below 1.02  $\mu$m is called NDR I, the wavelength region between 1.02  $\mu$m and 1.47 $\mu$m is called ADR I, and the wavelength region above 1.47 $\mu$m is called NDR II as labeled in Fig.~\ref{fig:WGADisp}. The circles mark the two pump wavelengths used in the simulations, 1.56 $\mu$m and 0.94 $\mu$m.  A wavelength independent loss of 1.35 dB/cm is used in the simulations~\cite{Krueckel2015a,Liu16SiRN}. 

The input pulse considered for the study has a sech$^2$ profile with time-domain full width at half maximum ($T_{FWHM}$) of 125 fs and a pulse energy of 82 pJ (58 W peak power) centered at 1.56 $\mu$m. The repetition rate used is 90 MHz, which is used to calculate the power spectral density. The spectral evolution along the length of the waveguide is plotted in Fig.~\ref{fig:WavSigPum15}(b). Initially the spectrum broadens from SPM in NDR II, and the shorter wavelength part of the spectrum leaks into the ADR I. The leaked pulse will in the ADR I develop into a soliton, which is subsequently repelled from the NDR II continuum. The repulsion can be explained by the cross-phase modulation (XPM) between the two pulses~\cite{Gu15Pob}, resulting in the frequency change from 
\begin{equation} \label{eq:XPM}
\partial\omega(T)\propto-\frac{\gamma_{1}L}{\pi}\frac{\partial}{\partial T}|A_{2}(L,T)|^{2},
\end{equation}
where $L$ is the interaction length for the two pulses, $\partial\omega(T)$ denotes change in instantaneous frequency of the first pulse, $\gamma_{1}$ is the nonlinear coefficient at the wavelength of the first pulse. $|A_{2}|^{2}$ is the power of the second pulse. As seen by the derivative in Eq.~(\ref{eq:XPM}), the change in the frequency caused by the interaction with the second pulse depends on the whether the interaction happens at the  leading or the trailing edge of the second pulse. By properly locating (delaying or advancing) the ADR pulse with respect to the NDR pulse, efficient blueshift or redshift of the ADR pulse is achieved. As seen in figure~\ref{fig:WavSigPum15}(d), the soliton overlaps in time domain with trailing edge of the NDR II pulse, therefore the soliton blueshifts because of XPM. As the soliton reaches the edge of the ADR I it sees a barrier~\cite{Skr03CSSFS}, and the soliton stops blueshifting. By this time, the initial pump pulse in the NDR II has temporally spread, greatly reducing the peak power. Therefore the XPM felt by the soliton is reduced. At around $z=22$ mm we see the formation of several DWs. Generation of DWs by a soliton is controlled by phase-matching conditions. A consequence of the specific phase-matching condition is that the DWs are generated outside the ADR. The soliton only phase matches one DW on each side of the ADR. Therefore, it is observed that DWs from the soliton alone do not explain all the DWs present in the spectrum. Here we explain the origin of the new DWs as non-degenerate FWM of the soliton and the pulse in the NDR II. The phase-matching conditions of both the degenerate and non-degenerate case is given by~\cite{Ylin04FWMDis}
\begin{equation}
\begin{split} \label{eq:PM1}
 \beta_{lin}(\omega_d)& =J[\beta_{lin}(\omega_p)-\beta_{sol}(\omega_p)]+\beta_{sol}(\omega_d);\\
   \textnormal{for}~J & =-1, 0, 1,
\end{split}
\end{equation}
where
\begin{equation}
\beta_{sol}(\omega)=\beta_{lin}(\omega_s)+\beta_1(\omega_s)[\omega-\omega_s]+q_{sol}.
\end{equation} 
Here $\beta_{lin}$ is the dispersion relation of the medium, $\omega_p$ denotes the center of the broadened pump spectrum that overlaps with the soliton in time, $\omega_s$ denotes the center frequency of the soliton, $\omega_d$ denotes the frequency of the DW and $q_{sol}$ is the soliton wavenumber. $q_{sol}$ has negligible contribution towards the calculated phase-matching wavelengths and is not included in the calculations.\begin{figure}[t!] 
\includegraphics[width=0.47\textwidth]{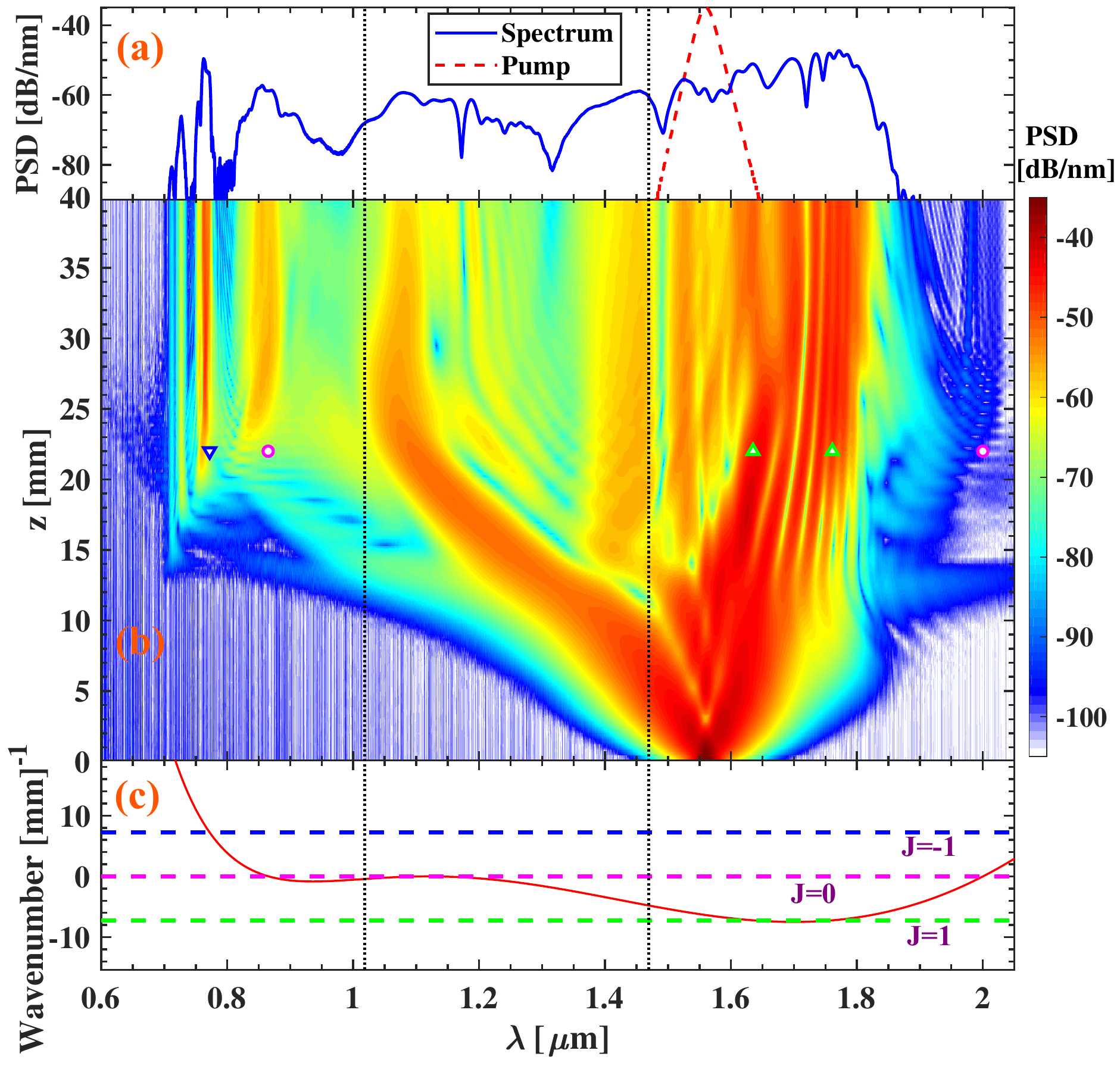} 
\includegraphics[width=0.47\textwidth]{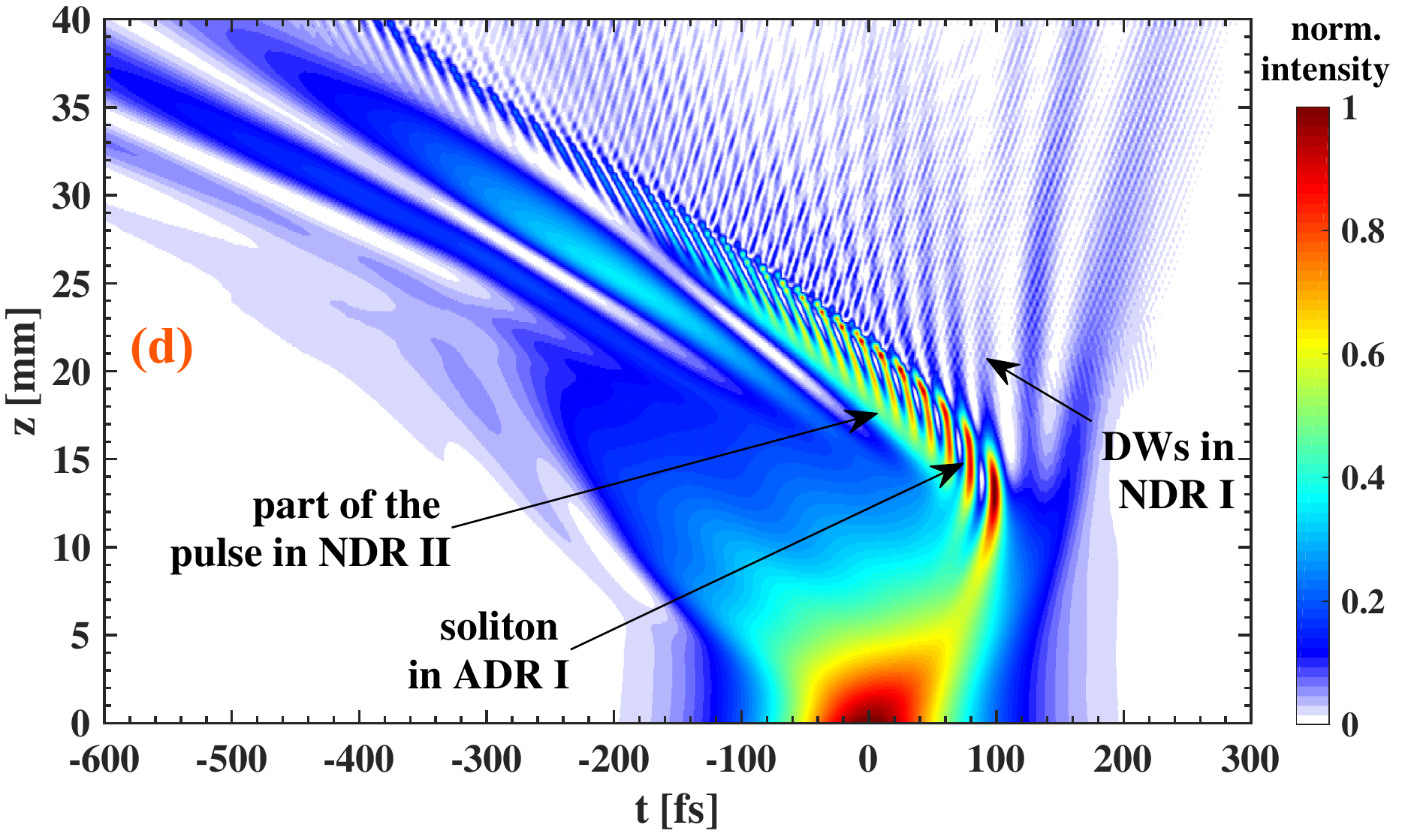}
\caption{Directional SCG in the SiRN waveguide pumped at 1.56 $\mu$m with 0.58 kW peak power. (a) The power-spectral density (PSD) at the beginning (dashed) and end of the SiRN waveguide (full). (b) The spectral evolution in the waveguide as a function of propagation distance, along with the phase matched wavelengths from Eq.~(\ref{eq:PM1}) after 22 mm of propagation in the waveguide, marked as downward pointing triangle ($J=-1$), empty circles ($J=0$), and triangles pointing upwards ($J=1$). (c) Dispersion relation of the waveguide in the soliton's group-velocity frame (full). The dashed line are the wavenumbers for $J=-1$, 0, 1 cases as labeled. The dotted lines in all the figures are the ZDWs. (d) Temporal evolution in the waveguide. Key spectral regions are marked. } 
\label{fig:WavSigPum15}
\end{figure}
\begin{figure}[htb!]
\includegraphics[width=0.47\textwidth]{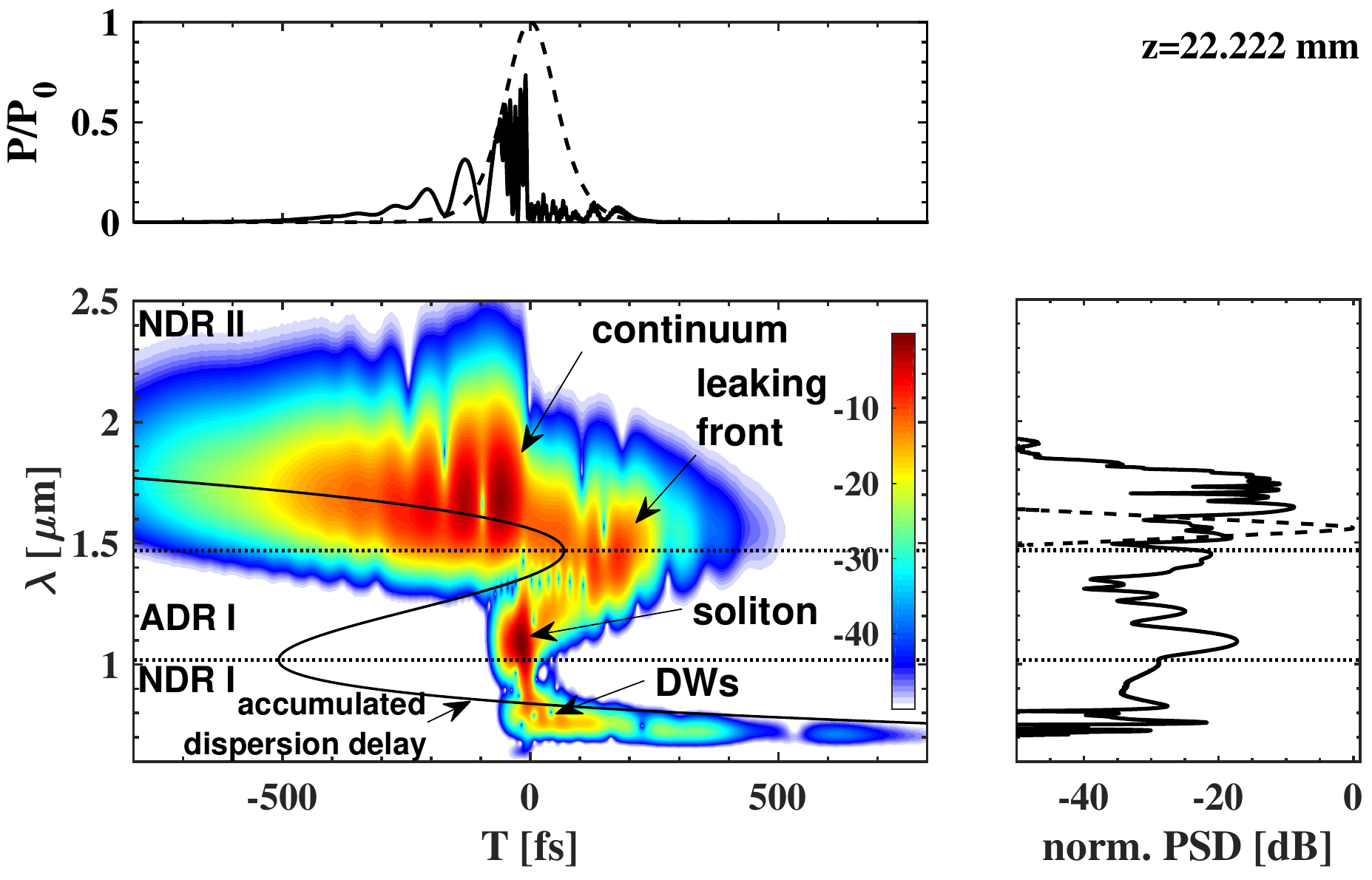}
\caption{Spectrogram at 22 mm calculated with a 16 fs gating pulse. The full black line shows the accumulated dispersion delay.}
\label{fig:Specgm15}
\end{figure}

The phase-matching points for the case at $z=22$ mm is graphically represented in Fig.~\ref{fig:WavSigPum15}(c). The points intersecting the dispersion relation of the waveguide in the soliton's group-velocity frame ($\beta_{lin}(\omega)-\beta_1(\omega_s)[\omega-\omega_s]$) with the $J=0$ line are the wavelengths at which the DWs directly from the soliton are generated. The empty circles  plotted in Fig.~\ref{fig:WavSigPum15}(b) marks the calculated phase-matching points where the DWs are expected to be generated. It can be seen that calculated points agree very well with the DWs that are generated. The phase-matching wavelength obtained from $J=-1$ condition is marked as triangles pointing down in Fig.~\ref{fig:WavSigPum15}(b). It is important to note the presence of another DW at a shorter wavelength in NDR I, next to the one generated by the soliton. This DW is the result of non-degenerate FWM of the soliton at 1.120 $\mu$m and the part of the pump that overlaps in time at 1.635 $\mu$m which leads to a phase-matching at 0.772 $\mu$m. It can be seen that $J=-1$ condition explains very well the origin of the DW next to the soliton's DW. The spectrum obtained after 40 mm is plotted in Fig.~\ref{fig:WavSigPum15}(a). The spectrum at the end of the waveguide spans from 0.750 to 1.85 $\mu$m. 290 nm broadening towards the longer wavelengths, and 790 nm towards the shorter wavelengths from the 1.56 $\mu$m pump. An enhanced spectral broadening towards the shorter wavelengths is observed. The directionality of the broadening is caused by the blueshifting of the wavelengths in the ADR and the emission of DW in the visible wavelengths.

In Fig.~\ref{fig:Specgm15} we show a calculated spectrogram at 22 mm propagation distance. At this stage the continuum in NDR II is well developed, and the blue trailing front has leaked into the ADR, generating the soliton. The soliton has experienced repulsion from the long-wavelength ZDW. Also, being on the other side of the ZDW its group-velocity has increased, so it is no longer overlapping with the tail of the continuum but is moving towards the center. At this stage, it has already generated several DWs in NDR I. In the plot we also show the accumulated dispersion delay $T(\omega,z)=[\beta_1(\omega)-\beta_1(\omega_0)]z$ where $\beta_1(\omega)$ is the frequency-dependent inverse group-velocity. This line indicates where the colors will be if they were all injected at $z=0$. Observing this line, it is clear that this allows the soliton to eventually interact with the entire continuum as it propagates along the waveguide: as the leaking front enters the ADR, the pulse will namely experience an acceleration and thereby be pulled through the continuum from the trailing to the leading edge. It is during this process that XPM between the continuum and the soliton will keep transferring energy to the DWs in NDR I.

\begin{figure}[tbp]
\includegraphics[width=0.47\textwidth]{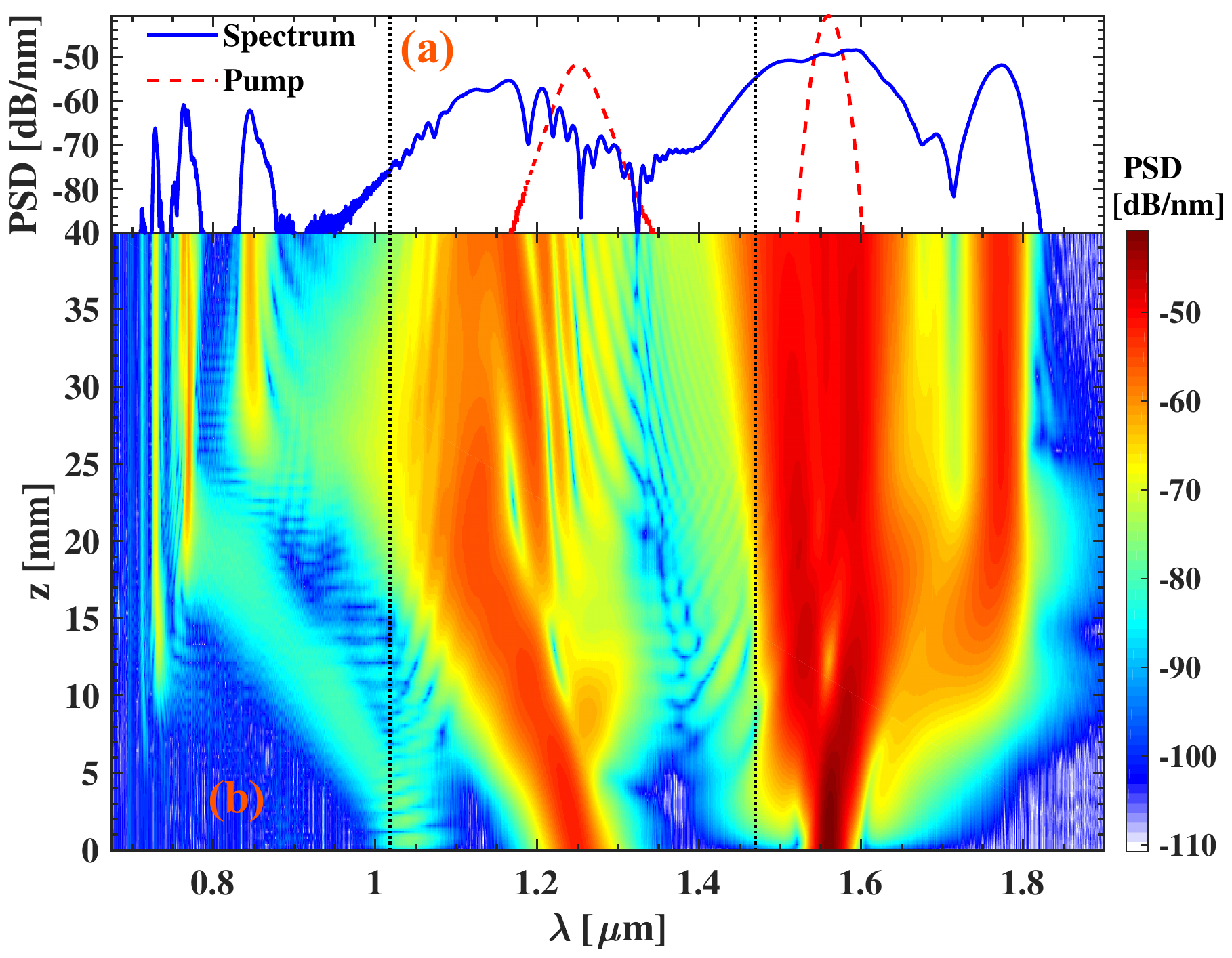}
\caption{Two-pulse model where the SiRN waveguide is pumped at 1.56 $\mu$m with 125 fs 140 W peak power and a soliton state is injected at 1.25 $\mu$m with 55 fs 40 W peak power. (a) Spectrum at the output of the SiRN waveguide (full) and the input (dashed). (b) Simulation of spectral evolution of a fundamental soliton colliding with a pulse in the NDR II. The vertical dotted lines mark the ZDWs in both the figures.}
\label{fig:WavTwoPum15}
\end{figure}

In the initial reporting on directional SCG~\cite{Oka17Coh}, the process was interpreted as a cascade DW phenomenon. Below it will be argued that exciting a soliton in the ADR is important for understanding the outcome of the phase-matching conditions. To confirm this, the launching of two pulses were implemented in our simulations. As a test case we modeled the collision of a strong pump in the NDR with an $N=1$ soliton in the ADR, where $N$ is the soliton order. Specifically, we simulated a 20 pJ 125 fs Gaussian pulse (140 W peak power) centered at 1.56 $\mu$m in NDR II colliding with a 2.5 pJ 55 fs soliton pulse (40 W peak power) centered at 1.25 $\mu$m in the ADR I. The low power is chosen for clarity of the interactions. The pulses are initially offset by 100 fs.

Figure~\ref{fig:WavTwoPum15}(b) shows the spectral evolution of the two pulses as they interact in the waveguide. First we see that the pulse in NDR II broadens to a continuum because of SPM, at the same time the soliton in the ADR I is blueshifted towards shorter wavelengths. This happens as the soliton is overlapping with the trailing edge of the pulse in NDR II. Therefore the XPM-induced spectral chirp is towards the short wavelengths. At around $z=12$ mm a clear peak emerges at 1.8 $\mu$m, which is the typical shoulder found after the onset of wave-breaking of the red pulse front of the NDR I non-solitonic continuum. This process occurs due to FWM between the SPM-broadened continuum and the undepleted pump that is temporally broadened due to dispersion. Later, at around $z=15$ mm, the soliton completely overlaps with the pump, and we observe the generation of a strong DW below 800 nm in NDR I. This is the non-degenerate FWM case with $J=-1$ between the soliton and the SPM-broadened pulse in the NDR II regime. The process quickly saturates when the latter depletes and all the photons are converted into the continuum~\cite{Finot08Ben}. At $z=25$ mm the soliton emits another dispersive wave caused by degenerate FWM. The overall evolution is similar to the evolution in the single-pulse simulation seen in Fig.~\ref{fig:WavSigPum15}. Figure~\ref{fig:WavTwoPum15}(a) shows the spectrum (full) after 40 mm of propagation in the waveguide, and qualitatively the generated spectrum is similar to the single-pump spectrum, both in terms of bandwidth and spectral shape with DWs in the 0.70-0.90 $\mu$m region and a bandwidth from 0.70 $\mu$m to 1.85 $\mu$m. This confirms our interpretation that the spectral broadening observed in the single-pulse simulation is well understood by considering the two-color interactions between a soliton in the ADR I and a non-solitonic continuum in NDR II. \begin{figure}
\includegraphics[width=0.47\textwidth]{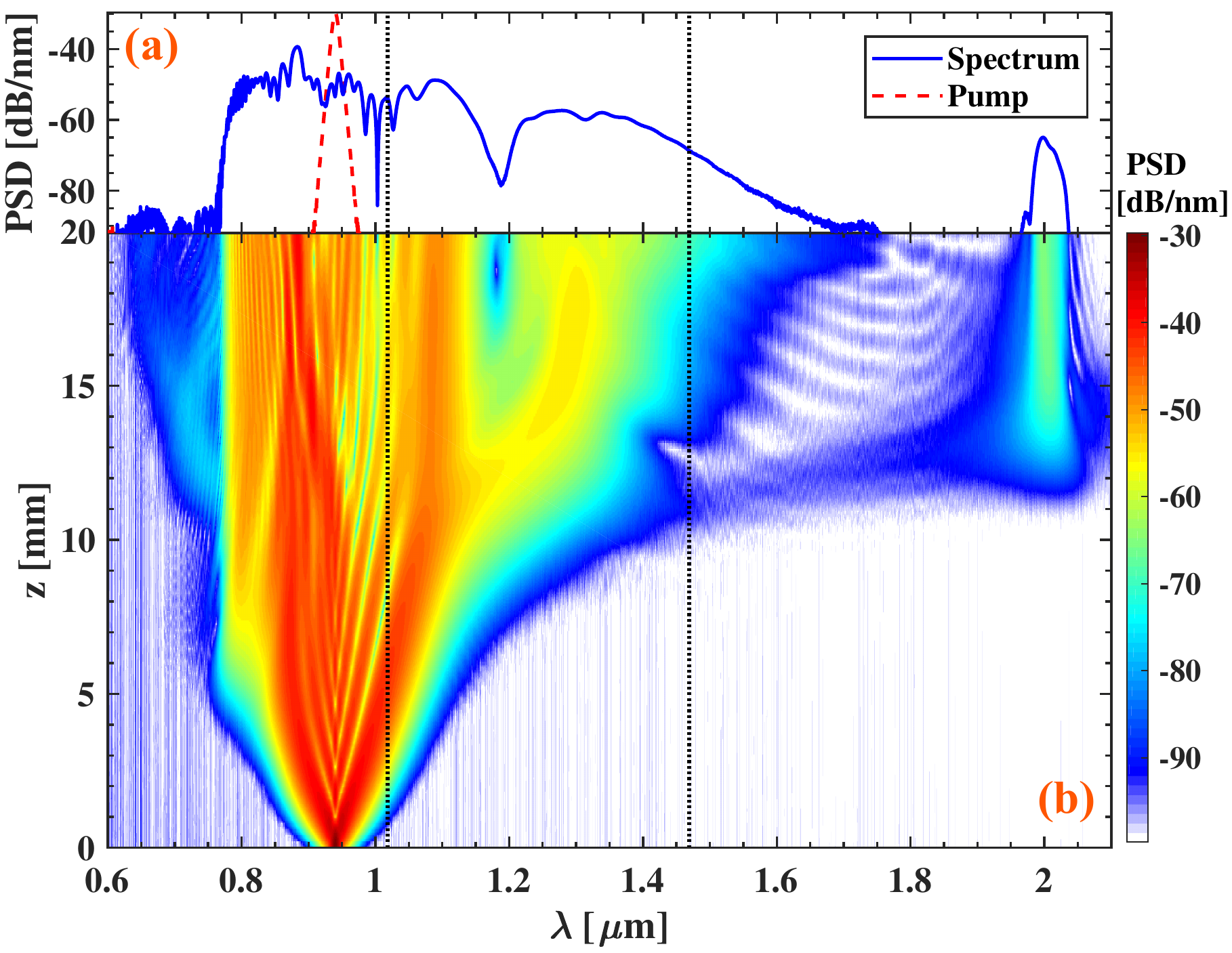}
\caption{Directional SCG in the SiRN waveguide pumped at 0.94 $\mu$m with 0.7 kW peak power. (a) Spectrum at the end of the waveguide (full) and the pump (dashed). (b) The spectrum evolution in the waveguide as a function of propagation distance for the waveguide pumped at 0.94 $\mu$m. The vertical dashed lines marks the ZDWs in both the figures. For this pump wavelength the nonlinear coefficient is $\gamma=7.2~\rm W^{-1}m^{-1}$.} \label{fig:WavSigPum9}
\end{figure}

We have observed that the 1.56 $\mu$m pulse propagating through the waveguide broadens towards the visible. To investigate the cause of directionality of the SCG, the wavelength of the single-pulse simulation is now changed to 0.94 $\mu$m, with $T_{FWHM}=125$ fs and 0.70 kW peak power. The spectral evolution and final spectrum can be seen in Fig.~\ref{fig:WavSigPum9}. Similar to the 1.56 $\mu$m case, the spectrum initially broadens because of SPM in NDR I. As it broadens it spreads into the ADR I. At $z=13$ mm the pulse in the ADR separates into two distinct peaks, and a DW with a wavelength of 2.0 $\mu$m is created by FWM between the ADR peak and the NDR pulse. The second peak is redshifted by XPM to 1.30 $\mu$m. The final spectrum has a bandwidth from 0.80 $\mu$m to 2.0 $\mu$m. This gives a broadening of 140 nm towards the shorter wavelengths and 1060 nm broadening towards the longer wavelength. Less power is transfered into the region between the IR dispersive wave and the $1.47 \mu$m ZDW. A similar spectral evolution as in the 1.56 $\mu$m case is observed, but this time it is directed towards the IR. From this it is concluded that the spectrum mainly broadens towards the direction of the ADR.
\subsection{Microstructured fiber}
MSFs have facilitated the study of SCG owing to the flexibility in obtaining a desired dispersion profile and ease in handling the fibers. SCG with fiber as a nonlinear medium has an additional advantage that it can be used to build an all-fiber compact set up with ease of alignment and portability. Due to their wide spread use in telecommunications, various silica fiber fabrication techniques are well understood. Silica fibers can have a low loss transmission window from 0.4 $\mu$m to 2.0 $\mu$m~\cite{Hill04TZD,Gri11LoWL}. Thus, silica MSF would be a natural choice for study of directional SC using a tabletop femtosecond fiber laser at 1.56 $\mu$m. To extend our understanding to that of fibers, we here analyze directional SC in a silica fiber with a dispersion profile similar to that studied in the SiRN waveguide.
\begin{figure}[t!]
\centering
\includegraphics[width=0.47\textwidth]{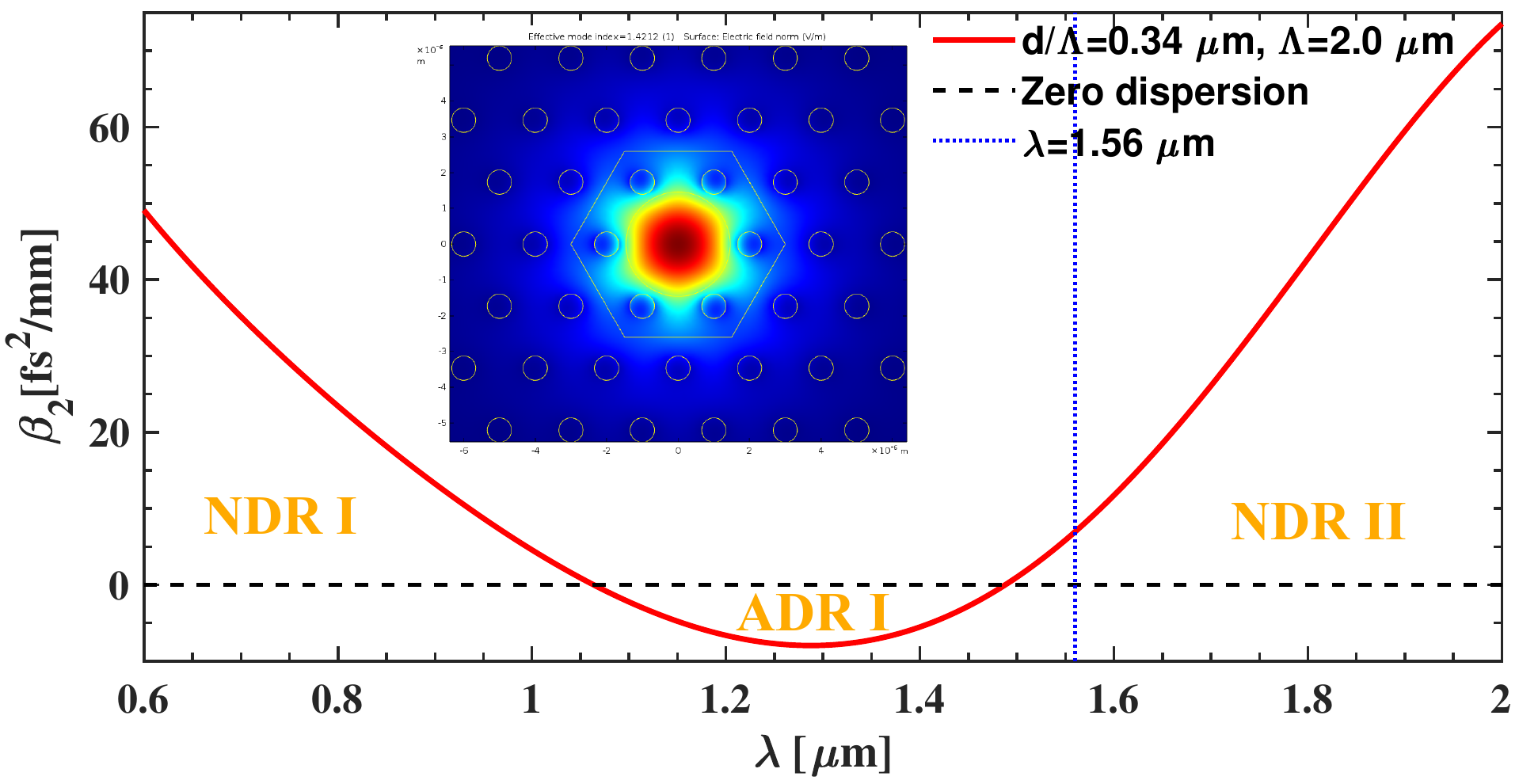}
\caption{GVD ($\beta_{2}$) of the MSF as a function of wavelength. The dashed horizontal line is the zero-dispersion and the dotted vertical line marks the pump wavelength at 1.56 $\mu$m. The inset represents the COMSOL simulation of the fiber and the mode obtained.} 
\label{fig:FibDisp}
\end{figure}

The pure silica fiber we consider has a pitch ($\Lambda$) of 2.0 $\mu$m and a hole diameter (d) of 0.680 $\mu$m. The modes in the fiber were found by using COMSOL Multiphysics. The numerically calculated dispersion of the fiber is plotted in Fig.~\ref{fig:FibDisp}; the fiber has ZDWs at 1.063 $\mu$m and 1.488 $\mu$m as shown in Fig.~\ref{fig:FibDisp}.  The fiber has a $\gamma$ of 7.65 W$^{-1}$km$^{-1}$ and an $A_{e f f}$ of 13.8 $\mu m^{2}$ at the pump wavelength of 1.56 $\mu$m. Both the material loss and the confinement loss of the fiber were included in the GNLSE. The material loss accounts for the loss arising from pure silica as described in Ref.~\cite{Mosl09The}. 

The GNLSE is solved in the frequency domain IP as done in the case of waveguide discussed previously. The complete $\beta(\omega)$ profile is included, and $f_{R}=0.18$, $\tau_{1}$=12.2 fs, and $\tau_{2}$=32 fs are used~\cite{Stol89Ram}. The same pump pulse configuration as in the SiRN case was used, except now the peak input power is 9 kW, corresponding to a pulse energy of 1.28 nJ. The spectrum obtained after 1 m of the fiber is plotted in Fig.~\ref{fig:Fib9kW}(a), while (b) shows the spectral evolution.

\begin{figure}[htb!]
\includegraphics[width=0.47\textwidth]{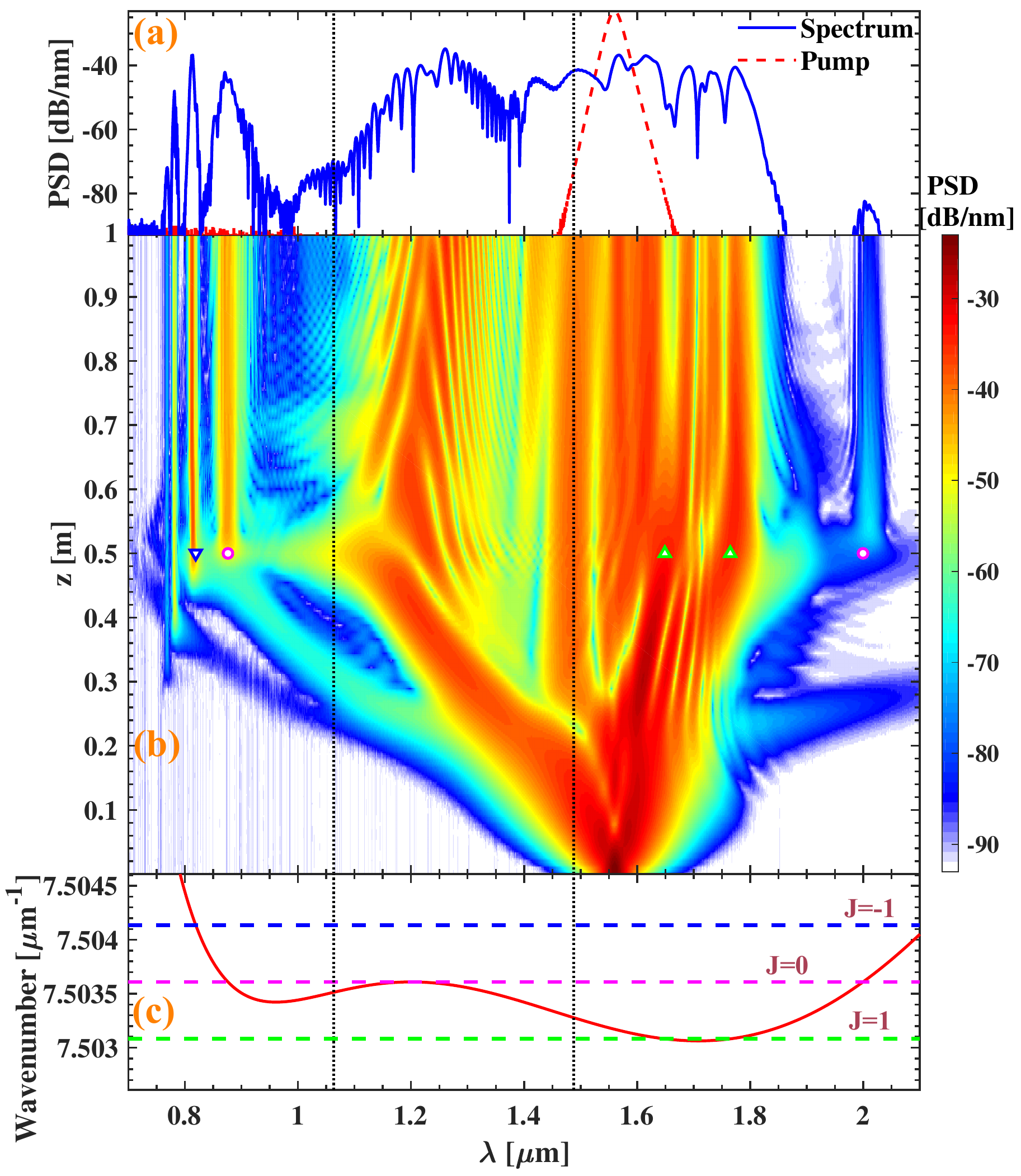}
\caption{Directional SCG in an MSF pumped at 1.56 $\mu$m with 9 kW peak power. (a) The PSD at the beginning (dashed) and end of the fiber (full). (b) Spectral evolution along the length of the fiber. The empty circle at 0.876 $\mu$m and 1.999 $\mu$m are the calculated wavelength at which the DWs are generated from phase-matching condition for the soliton at 1.20 $\mu$m alone after 0.5 m of propagation in the fiber. The downward pointing triangle at 0.819 $\mu$m is the calculated wavelength with $J=-1$ condition in Eq.~(\ref{eq:PM1}), at which DW are generated from the non-degenerate FWM of the part of the pump at 1.649 $\mu$m in NDR II and the soliton after 0.5 m of the fiber. The triangles pointing upwards at 1.649 $\mu$m and 1.765 $\mu$m are the calculated wavelength with $J=1$ condition in Eq.~\ref{eq:PM1}, at which DW are generated after 0.5 m of the fiber. (c) Dispersion relation of the waveguide in the soliton's group-velocity frame after 0.5 m of propagation in the fiber (full). The dashed lines are the wavenumbers for $J=-1$, 0, 1 cases as labeled. The intersection points between the full line and the dashed lines are calculated phase-matching wavelengths for the generation of DWs. The dotted vertical lines in all three figures are the ZDWs.} 
\label{fig:Fib9kW}
\end{figure}

Figure~\ref{fig:Fib9kW}(c) shows the plot of dispersion relation of the fiber in the soliton's group-velocity frame. The center wavelength of the soliton is 1.20 $\mu$m at 0.5 m of the fiber. The distinct features of the spectral evolution shown in Fig.~\ref{fig:Fib9kW}(b) can be explained as follows. The pump is in the NDR II, and the spectrum initially broadens from SPM. A part of the spectrum spreads into the ADR I and develops into a soliton. Initially, the soliton blueshifts from XPM induced from a part of the pump in NDR II that overlaps in time. The soliton shifts towards the shorter ZDW. After around 0.5 m of propagation, four DWs are generated: two in NDR I and two in NDR II. Further propagation spreads out the power in the NDR II, and the soliton does not experience XPM from the pulse in the NDR II region anymore. The soliton, which is at the edge of the ADR I, begins to redshift from Raman-induced self-frequency shift. As in the case of the waveguide the DWs from the soliton alone does not explain all the DWs observed in the spectral evolution at 0.5 m of propagation in the fiber, as can be seen in Fig.~\ref{fig:Fib9kW}(b). The new DWs are well explained by considering them as DWs generated from the non-degenerate FWM of the part of the pump at 1.649 $\mu$m in NDR II that overlaps in time with the soliton at 1.20 $\mu$m. This is shown in Eq.~(\ref{eq:PM1}) and explained in Fig.~\ref{fig:Fib9kW}. It is interesting to note that the band of wavelength generated at around 1.75 $\mu$m can be explained as a dispersive wave generated from the non-degenerate FWM of the soliton at 1.20 $\mu$m and the part of the pump that overlaps in time at 1.649 $\mu$m.
\begin{figure}[htbp!]
\includegraphics[width=0.47\textwidth]{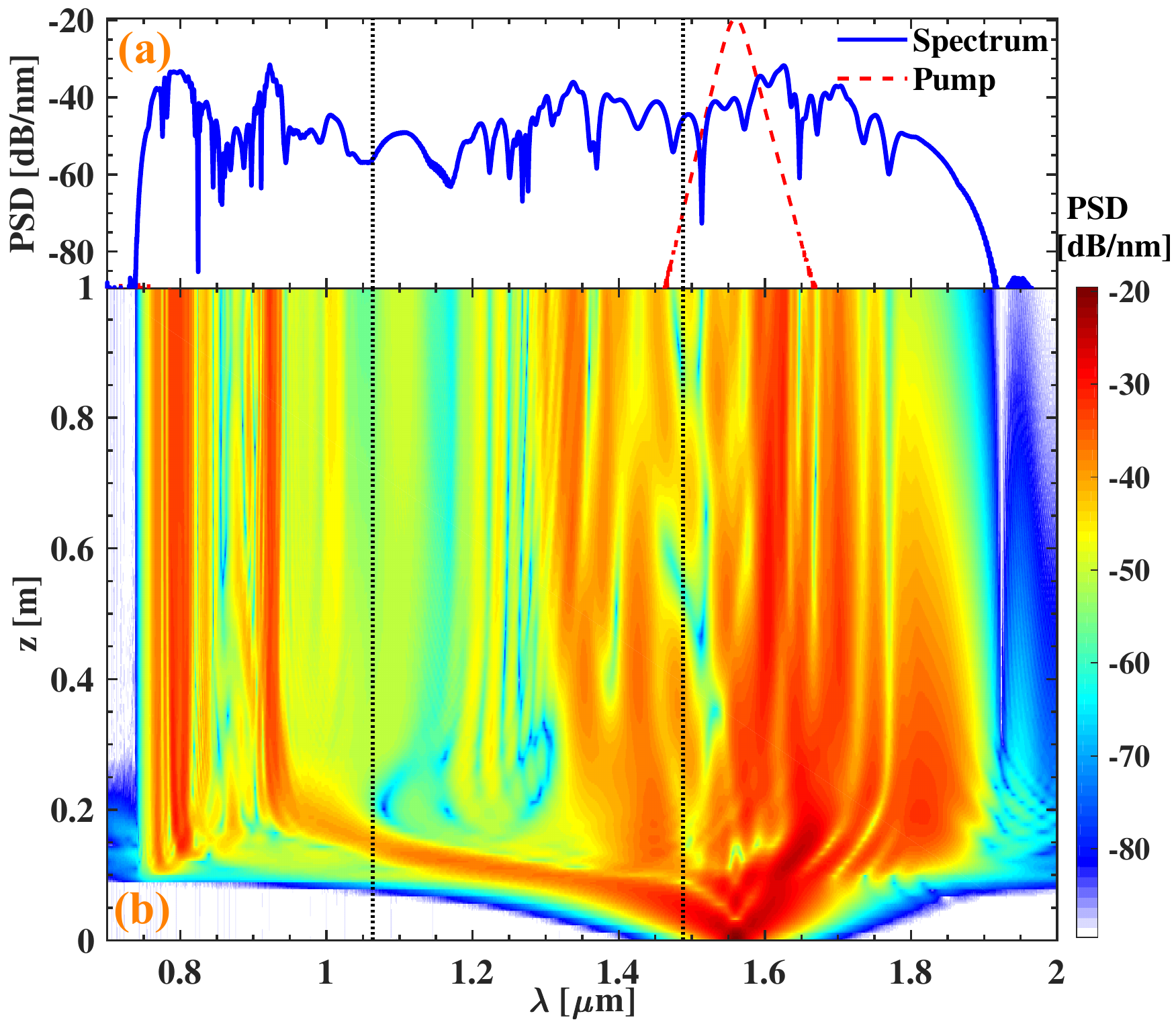}
\caption{Directional SCG in an MSF pumped at 1.56 $\mu$m with 20 kW peak power. (a) The PSD at the beginning (dashed) and end of the MSF (full). (b) Spectral evolution along 1 m of the fiber. The vertical dotted lines in both the figures are the ZDWs.}
\label{fig:Fib20kW}
\end{figure}
To further study the effect of XPM induced by the part of pump pulse in the NDR II overlapping in time with the soliton, the peak power is increased to 20 kW with all the other parameters remaining the same. The spectrum obtained after propagating 1 m of fiber is plotted in Fig.~\ref{fig:Fib20kW}(a), and the spectral evolution is shown  Fig.~\ref{fig:Fib20kW}(b). It can be observed that similar to the 9 kW case, the spectrum initially broadens from SPM in NDR II and a part of the spectrum spreads in to the ADR I. There it forms a soliton, which is repulsed by XPM from the NDR II continuum, resulting in a soliton blueshift just like in the previous cases. The spectrogram at $z=0.122$ m in the fiber is shown  in Fig.~\ref{fig:Sgm20kW}(a). It can be seen from the Fig.~\ref{fig:Sgm20kW}(a) that the co-propagating broad pulse in NDR II overlaps in time with the soliton in ADR I and that it emits DWs on the short wavelength side. The spectrogram calculated at z=0.122 m is bandpass filtered from 1.1 to 1.5 $\mu$m to see the pulse in ADR I. The temporal profile and the spectrogram in Fig.~\ref{fig:Sgm20kW}(c) shows that the pulse at 1.16 $\mu$m is indeed a soliton. Interestingly, as the soliton encounters the ZDW barrier, it does not recoil as seen in the previous cases, but continues through. This can be qualitatively explained as follows. Compared to the 9 kW case, the repulsion velocity in this case is much higher, as a higher peak power (20 kW) is launched into the fiber. This makes the pulse in NDR II much stronger than in the 9 kW  case. Thus, the XPM experienced by the soliton is much greater, resulting in a higher repulsion velocity towards the shorter wavelengths. The spectrogram in Fig.~\ref{fig:Sgm20kW}(b) shows the soliton being ``pushed'' into the NDR I. Once in NDR I, the pulse continues to blueshift until the power in the NDR II spreads out and there is no more XPM experienced by the pulse. As more of the spectrum from the NDR II spreads into the ADR I it develops into solitons and gradually redshift from Raman-induced self-frequency shift.
\begin{figure}[htbp!]
\includegraphics[width=0.47\textwidth]{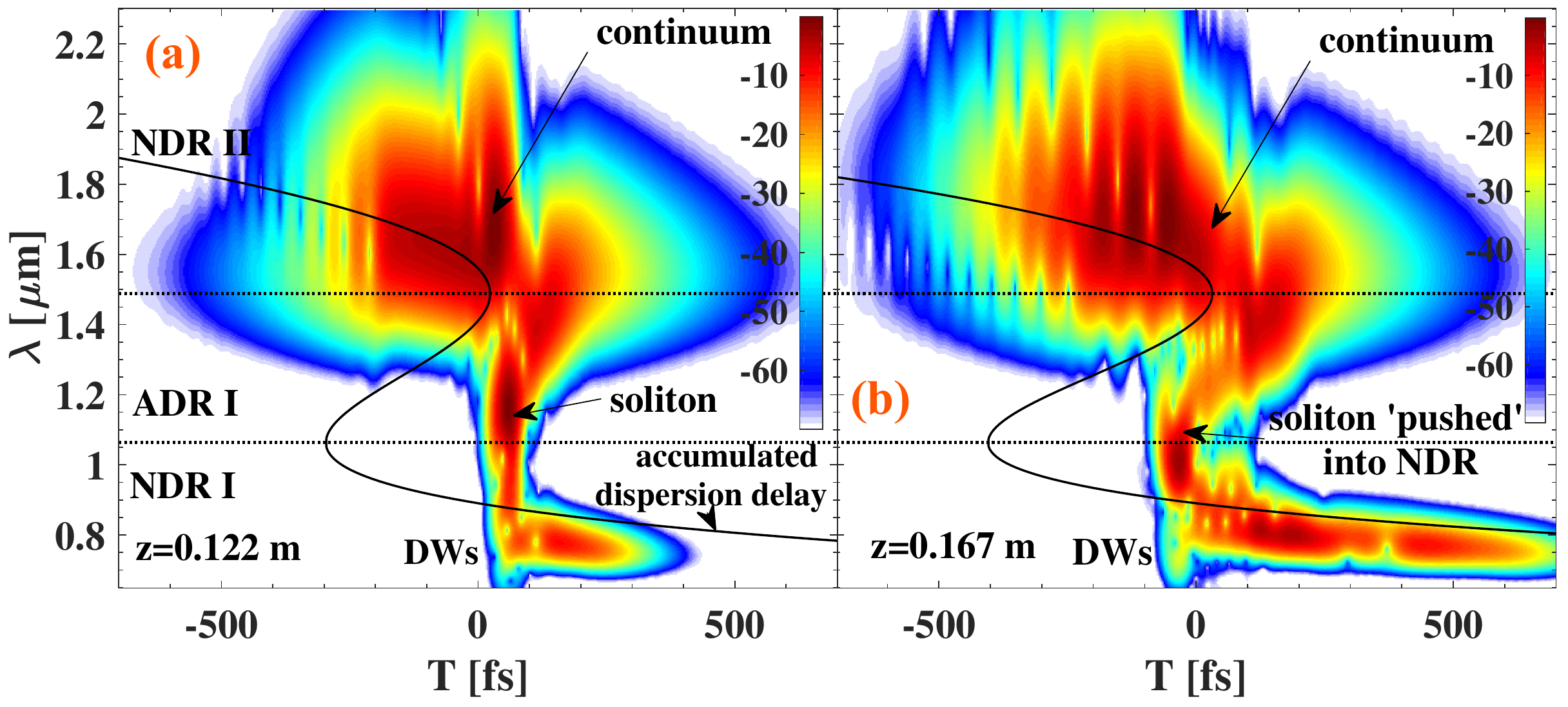}
\includegraphics[width=0.47\textwidth]{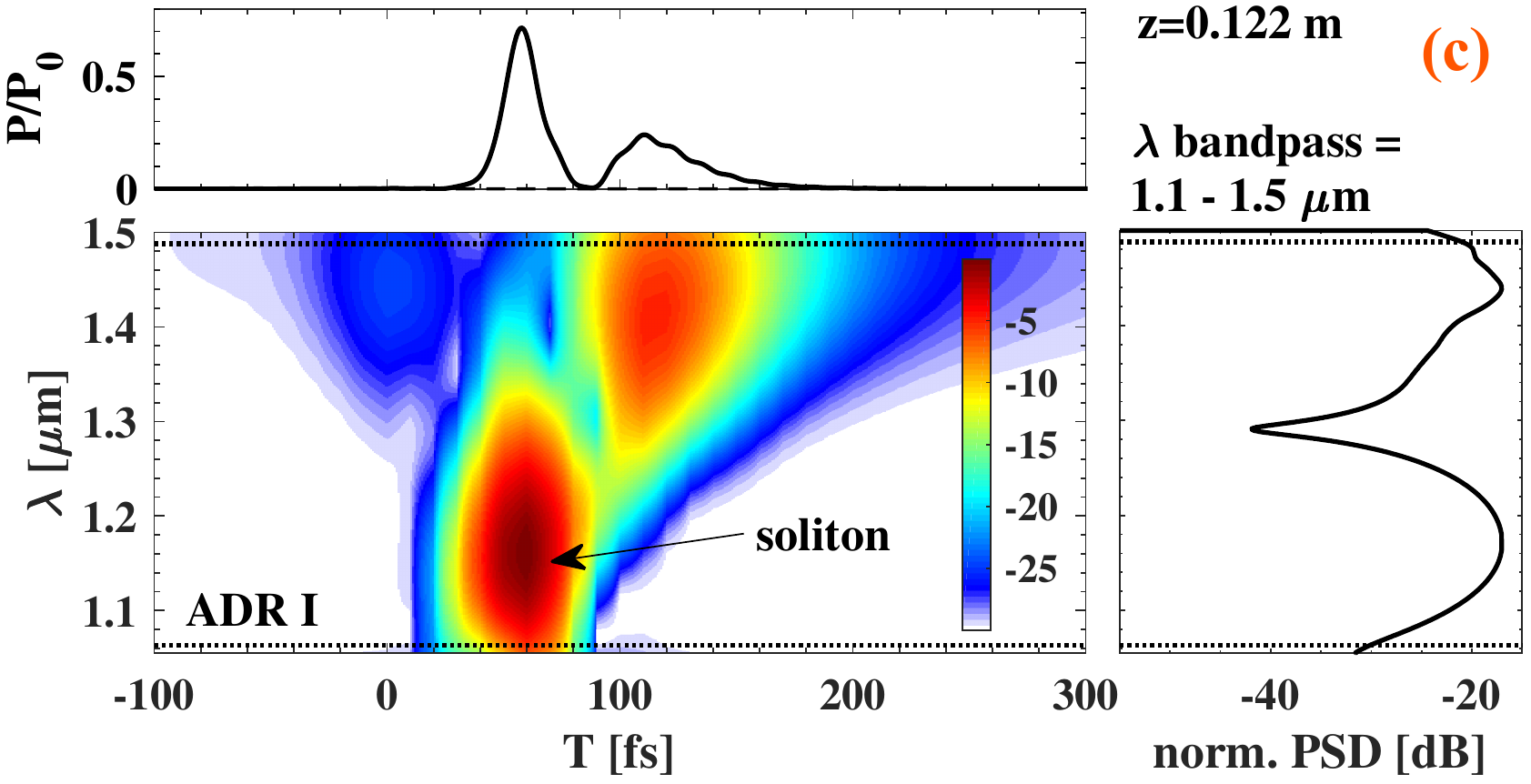}
\caption{Spectrograms for an input peak power of 20 kW calculated with a gating function of 20 fs: (a) at z=0.122 m, showing the soliton in the ADR I and the co-propagating pulse in NDR II; (b) at z=0.167 m, shows the soliton being pushed into the NDR I. (c) Spectrogram numerically bandpass filtered form 1.1 to 1.5 $\mu$m to show the spectral and temporal profile in the ADR I at z=0.122 m.}
\label{fig:Sgm20kW}
\end{figure}

A future challenge is to understand why the soliton passes the ZDW barrier and how come it does not recoil; we did confirm that it is a soliton in the ADR that we see in the spectrum as shown in Fig.~\ref{fig:Sgm20kW}(c), but as soon as it encounters the ZDW it passes through and starts developing a chirp due to the normal-dispersion. A possible explanation is that due to the higher repulsion velocity the soliton encounters the ZDW at a more rapid pace than the previous cases, and while it starts to shed energy into the DW just above 0.8 $\mu$m in NDR I, the accompanying spectral recoil is not enough to stop the soliton from passing the ZDW.
\section{Conclusion}
We have investigated a directional SCG scheme where the waveguide has an anomalous-dispersion region (ADR) surrounded by two normal-dispersion regions (NDRs). When pumping in one of the NDRs, the SC broadening is directed mainly through the ADR and towards the other NDR. This directional broadening is shown to be caused by a 3-step process. First, since the pump is in the NDR, it broadens due to SPM into a typical normal-dispersion continuum, where each pulse front broadens both spectrally and temporally. The second step is when one of the pulse fronts reaches the ZDW. It leaks into the ADR, and a soliton forms in the ADR as the anomalous-dispersion compensates for the accumulated nonlinear phase. The soliton is repulsed spectrally from the continuum in the NDR due to XPM between them. The third step is formation of DWs in the other NDR, which occurs either as direct FWM with the soliton or as non-degenerate FWM between the soliton and the continuum in the other NDR (i.e. an XPM-initiated process). In this process the soliton repulsion is stopped, and it recoils at the second ZDW. 

The directional SCG process was investigated through numerical simulations in a SiRN waveguide with the ADR between 1.02 and 1.47 $\mu$m. We found that when pumping at 1.56 $\mu$m with a fs laser the SC was directed towards  shorter wavelengths, while when using a 0.94 $\mu$m fs pump the SC was directed towards longer wavelengths. To show the generality of the scheme, we also showed simulations of a MSF geometry based on conventional silica fiber technology, where the hole pitch and diameter are selected to give a similar dispersion landscape. Qualitatively identical results were obtained concerning the nonlinear dynamics, with the main differences stemming from the different dispersion profiles. Finally, we also investigated using a higher pump power and found that the soliton gained a stronger repulsion velocity and higher peak power; and instead of recoiling at the other ZDW, it simply crosses the ZDW into the NDR. It remains to be discovered exactly why this happens, it is believed to be caused by the high repulsion velocity preventing the soliton to emit enough dispersive waves for the spectral recoil to overcome the XPM repulsion.

One main conclusion is that directional SCG is observable both in fibers and integrated photonics waveguides. This is welcome since these nonlinear platforms offer complementary solutions (transverse size, length, nonlinearity, dispersion engineering) for different communities. By choosing a suitable dispersion profile and using the required pump power, this technique could be used as a method to efficiently transfer pump power into a given wavelength region of interest. One exciting prospect could be to design the dispersion such that the ADR lies above the 1.56 $\mu$m pump wavelength, either in a SiRN waveguide or a soft-glass MSF. The SC would then directionally broaden towards the mid-IR where both of these waveguides have transmission.
 
\setlength\parindent{0pt}\vspace{3 ex}
{\large\sffamily\bfseries Funding.} Horizon 2020 Framework Programme (H2020) Research and Innovation Program under Marie Sk\l{}odowska-Curie (722380) (SUPUVIR).

\setlength\parindent{0pt}\vspace{3 ex}
{\large\sffamily\bfseries Acknowledgment.}
Victor Torres-Company and Zhichao Ye are acknowledged for data and design of the silicon nitride waveguide. S. C. acknowledges funding through the Nordic Five Tech alliance. 

\vspace{2ex}
\balance
\bibliography{DirSup}
\end{document}